\documentclass[usenatbib]{mn2e}
\usepackage{apjfonts,amsfonts,amsmath,amssymb,bm,ctable,verbatim}

\newcommand{\FIREurl}{\href{http://fire.northwestern.edu}{\url{http://fire.northwestern.edu}}}
\newcommand{\gizmourl}{\href{http://www.tapir.caltech.edu/~phopkins/Site/GIZMO.html}{\url{http://www.tapir.caltech.edu/~phopkins/Site/GIZMO.html}}}

\newcommand{\acknowledgments}[1]{\begin{small}\section*{Acknowledgments}\end{small}{\noindent #1}\vspace{5pt}}

\newcommand{\beq}{\begin{equation}}
\newcommand{\eeq}{\end{equation}}

\title[Seeds Don't Sink]{Seeds Don't Sink: Even Massive Black Hole ``Seeds'' Cannot Migrate to Galaxy Centers Efficiently}

\author[Ma et al.]{
\parbox[t]{\textwidth}{Linhao Ma$^{1,}$\thanks{E-mail: lma3@caltech.edu}, Philip F.~Hopkins$^1$, Xiangcheng Ma$^2$, Daniel Anglés-Alcázar$^{\,3,4}$, Claude-André Faucher-Giguère$^5$ and Luke Zoltan Kelley$^5$} \vspace*{4pt} \\
$^1$ TAPIR, Mailcode 350-17, California Institute of Technology, Pasadena, CA 91125, USA\\
$^2$ Department of Astronomy and Theoretical Astrophysics Center, University of California Berkeley, Berkeley, CA 94720\\
$^3$ Department of Physics, University of Connecticut, 196 Auditorium Road, U-3046, Storrs, CT 06269-3046, USA\\
$^4$ Center for Computational Astrophysics, Flatiron Institute, New York, NY 10011, USA\\
$^5$ CIERA and Department of Physics \& Astronomy, Northwestern University, Evanston, IL 60208, USA
}

\date{}
\begin{document}

\maketitle
\mathchardef\mhyphen="2D 

\begin{abstract}
Possible formation scenarios of supermassive black holes in the early universe include rapid growth from less massive seed black holes (BHs) via super-Eddington accretion or runaway mergers, yet both of these scenarios would require seed BHs to efficiently sink to and be trapped in the galactic center via dynamical friction. This may not be true for their complicated dynamics in clumpy high-$z$ galaxies. In this work we study this ``sinking problem'' with state-of-the-art high-resolution cosmological simulations, combined with both direct $N$-body integration of seed BH trajectories and post-processing of randomly generated test particles with a newly developed dynamical friction estimator. We find that seed BHs less massive than $10^8\,M_\odot$ (i.e., all but the already-supermassive seeds) cannot efficiently sink in typical high-$z$ galaxies. We also discuss two possible solutions: dramatically increasing the number of seeds such that one seed can end up trapped in the galactic center by chance, or seed BHs being embedded in dense structures (e.g. star clusters) with effective masses above the mass threshold. We discuss the limitations of both solutions.
\end{abstract}


\section{Introduction}
\label{sec:intro}

Supermassive black holes (SMBHs) are of crucial importance in understanding galaxy formation and evolution. Observations of high-redshift quasars have confirmed the existence of SMBHs in the first billion years after the Big Bang (\citealt{Fan2001,Fan2003,Lawrence2007,Willott2007,Morganson2012}, see Figure 1 of \citealt{Inayoshi2019} for a summary of observations). One of the long standing problems with models of SMBHs regards how they could possibly grow to such an enormous mass in a relatively short time period \citep{Turner1991}. Recent discoveries have found both extremely massive SMBHs in the early universe (e.g. SDSS J010013.02+280225.8 as a $1.2\times 10^{10}\,M_\odot$ SMBH at $z=6.3$, see \citealt{Wu2015}) and massive SMBHs in the extremely early universe (e.g. ULAS J1342+0928 as a $7.8\times10^8\,M_\odot$ SMBH at $z=7.54$, see \citealt{Banados2018}). Continued discoveries of SMBHs at higher redshifts and masses naturally makes the problem even more intriguing \citep{Haiman2001,Natarajan2014}.

The existence of such massive black holes (BHs) at such early times poses many unsolved theoretical challenges. The most well-known is the ``timescale problem'': if seeds begin life as much less massive BHs, they would have to accrete at $\sim 100\%$ of the Eddington limit, for $\gtrsim100\%$ of the age of the universe to reach their observed masses at $z>7$. But observations at all lower redshifts, and theoretical estimates of the effect of SNe and BH feedback and BH dynamics all argue for much lower duty cycles (see,. e.g., \citealt{Johnson2007,Whalen2008,Alvarez2009,Milosavljevic2009,Habouzit2017}. An obvious possible solution is to form more massive seeds: it has been proposed that primordial gas at high-$z$ could experience inefficient cooling and fragmentation, producing massive Population III stars \citep{Bromm2004} which could collapse to BH seeds as large as $\sim 100\,M_{\odot}$ \citep[e.g.][]{madau2001,Li2007,volonteri2012,Hirano2014} or even hyper-massive quasi-stars which could leave seeds as large as $\sim 10^{4}-10^{5}\,M_{\odot}$ (e.g. \citealt{Bromm2003,Hosokawa2012,Hosakawa2013,Hirano2017,Inayoshi2018}), or directly collapsing to BHs as massive as $10^5\,M_\odot$ \citep{Ludato2006,Ludato2007}. Yet several authors argue that this requires vanishingly improbable conditions (see, e.g. \citealt{moran2018} and discussions in \S~5.2 and \S~5.3 from \citealt{Inayoshi2019}). But even these most-optimistic models only reduce the timescales by a logarithmic factor (as timescales scale as $\log (M_\mathrm{SMBH}/M_\mathrm{seed})$): even in these models, a phase of highly super-Eddington accretion -- either resulting from runaway gas capture in high-gas-density regions (e.g. \citealt{ Madau2014,Alexander2014,Lupi2016,Pezzulli2016,Regan2019,Natarajan2021}), or runaway mergers of massive stars (e.g. \citealt{Portegies-Zwart2004,Devecchi2009,Katz2015,Reinoso2018}) or of other seed BHs (e.g. \citealt{Davies2011,Lupi2014}) at the center of a common potential minimum undergoing dynamical relaxation -- is likely needed to explain SMBHs at $z>7$ \citep{Haiman2004,Kawashima2012,Pacucci2015,Inayoshi2016,Ryu2016,Takeo2019}.

However, in the past two decades, many independent studies (e.g. , focused on galaxy mergers \citep{Governato1994,Volonteri2005,Callegari2009,Bellovary2010,2018ApJ...857L..22T}, dwarf galaxy evolution \citep{Tamfal2018,Bellovary2019,2020MNRAS.495L..12B} and/or BH growth/dynamics \citep{Callegari2011,Volonteri2016,Volonteri2020,daa:BHs.on.FIRE,2017MNRAS.469..295B,2019MNRAS.486..101P,Barausse2020,Catmabacak2020}) have pointed out that {\em all} of these models face a different and potentially even more severe challenge: what we refer to as the ``sinking problem.'' In brief: essentially all of the rapid/efficient accretion models require that BHs sink ``efficiently'' and remain tightly bound to the galaxy center or potential minimum, where densities are on average highest. This usually requires a well-defined and stable dense central region in a relatively massive galaxy at lower redshift ($z\lesssim 4$) \citep{Tremmel2017,Tremmel2019,Ricarte2019}, but it may not be possible {\em dynamically} for even ``high'' mass seeds in realistic turbulent, clumpy, high-redshift ($z\gtrsim7$) galaxies which undergo frequent dynamical perturbations (from e.g.\ mergers and ``bursty'' star formation and stellar feedback) and lack such central regions, especially in the short timescales available. Observationally, SMBHs are seen in the galactic center for most massive quasi-stellar objects (QSOs) (including those at high-$z$ where imaging is possible, e.g. \citealt{venemans2017,banados2019,decarli2019,Novak2019,wang2019}, and almost {\em all} massive galaxies comparable to QSO hosts at low redshifts, see e.g. \citealt{Ferrarese2000,gebhardt2000,tremaine2002,Graham2011,beifiori2012}). But in spatially-resolvable low-$z$ dwarf galaxies where star formation is known to be ``bursty'' \citep{Weisz2014,sparre2017,FaucherGiguere2018,Velazquez2020} and there is no well-defined dynamical center (see e.g. \citealt{Kallivayalil2013}), AGNs are extremely rare and those identified are randomly-scattered in position around the galaxy \citep{Reines2020,Mezcua2020}. As numerical simulations of high-$z$ galaxies have improved in both numerical resolution and incorporating the physics of star formation and stellar feedback in a turbulent, multi-phase ISM \citep{anglez-alcazar2014,daa:BHs.on.FIRE,Kretschmer2020}, most models have converged toward the prediction that high-$z$ galaxies are clumpy, bursty, chaotic, and dynamically-unrelaxed systems (even more so than most local dwarfs, e.g. \citealt{Tamfal2018,Muratov2015,Oklopcic2017,xiangchengma2018,kim2019,meng2020}), in agreement with deep observations with the {\em Hubble Space Telescope} (HST) \citep{elmegreen2007,Overzier2010,swinbank2010}. Although there is some evidence for rotation in some hosts as noted by, e.g. \cite{Decarli2018,Venemans2019}, they usually exhibit very large dispersion with $\sigma\sim v$, consistent with the simulations analyzed in \cite{Ma2017}, which does not challenge the conclusion. But in almost all models for rapid BH growth at near-Eddington or super-Eddington rates at $z\gtrsim7$, the most optimistic assumption possible is usually made: namely that the BH remains ``anchored'' to the local potential minimum at the center of some well-ordered galaxy (e.g. \citealt{Li2007}). To accrete gas, the BH must first capture it from the surroundings, and dimensional estimates for the ``capture rate'' drop highly super-linearly and extremely rapidly if the BH or background medium are moving relative to one another and/or if the BH lies outside of the galactic density maximum \citep{Hoyle1939}. Models like runaway stellar mergers or BH-BH seed mergers for rapid growth fundamentally {\em depend} on the idea that both the ``main seed'' BH and all other stars/seeds are anchored to and sinking rapidly towards a common dynamical center \citep{PortegiesZwart2002,Gurkan2004,shi2020,Gonzalez2020}.

Historically, the ``sinking'' of BH seeds in high-$z$ galaxies has largely been studied by assuming (1) seeds form at the centers of their proto-galaxies (rather than where stars form or at local density maxima), (2) galaxies are smooth objects with well-defined dynamical centers and centrally-peaked density profiles (i.e.\ bulge+disk or isothermal sphere models, rather than messy, non-relaxed systems), and (3) that BH and merging galaxy orbits decay according to dynamical friction (DF), which is a statistical accumulative effect caused by successive two-body gravity encounters, effectively acting like a ``drag force'' proportional to the BH/merging galaxy mass, in which the traditional \citet{chandrasekhar1943} (C43) DF formula (assuming a homogeneous, infinite, idealized background medium) is applied. In this paper, we therefore revisit the ``sinking'' and ``retention'' problems for seed BHs in early galaxies. We use high-resolution cosmological simulations which include the crucial physics described above, combined with both direct (``live'') $N$-body integration of seed BH trajectories and semi-analytic orbit integration in post processing, to follow a wide range of possible BH seed populations with different formation properties and locations. In post processing, we apply a modified DF estimator developed in a companion paper (Ma et al. in prep.), which is more flexible, accurate, and computationally efficient. In \S~\ref{sec:methods} we describe our numerical simulations and the semi-analytic post-processing method.

The plan of this paper is as follows: in \S~\ref{sec:results} we present the results from simulations and semi-analytical integration of sample orbits, and show that seed BHs are generally not able to sink efficiently or be retained even at high seed masses. In \S~\ref{sec:discussion}, we discuss possible solutions to this problem, but also use our simulations to highlight how these solutions encounter still other problems. We summarize in \S~\ref{sec:conclusions}.

Throughout, we assume a standard flat $\mathrm{\Lambda CDM}$ cosmology with $\Omega_\mathrm{m} = 0.31$, $\Omega_\Lambda = 1-\Omega_\mathrm{m}$, $\Omega_\mathrm{b}= 0.046$, and $H_0 = 68\,\mathrm{km\,s}^{-1}\mathrm{Mpc}^{-1}$ (e.g. \citealt{Planck2020}).

\section{Methods}
\label{sec:methods}

\subsection{Direct Simulations}
\label{sec:methods:sims}

\begin{figure*}
    \centering
    \includegraphics[width=\textwidth]{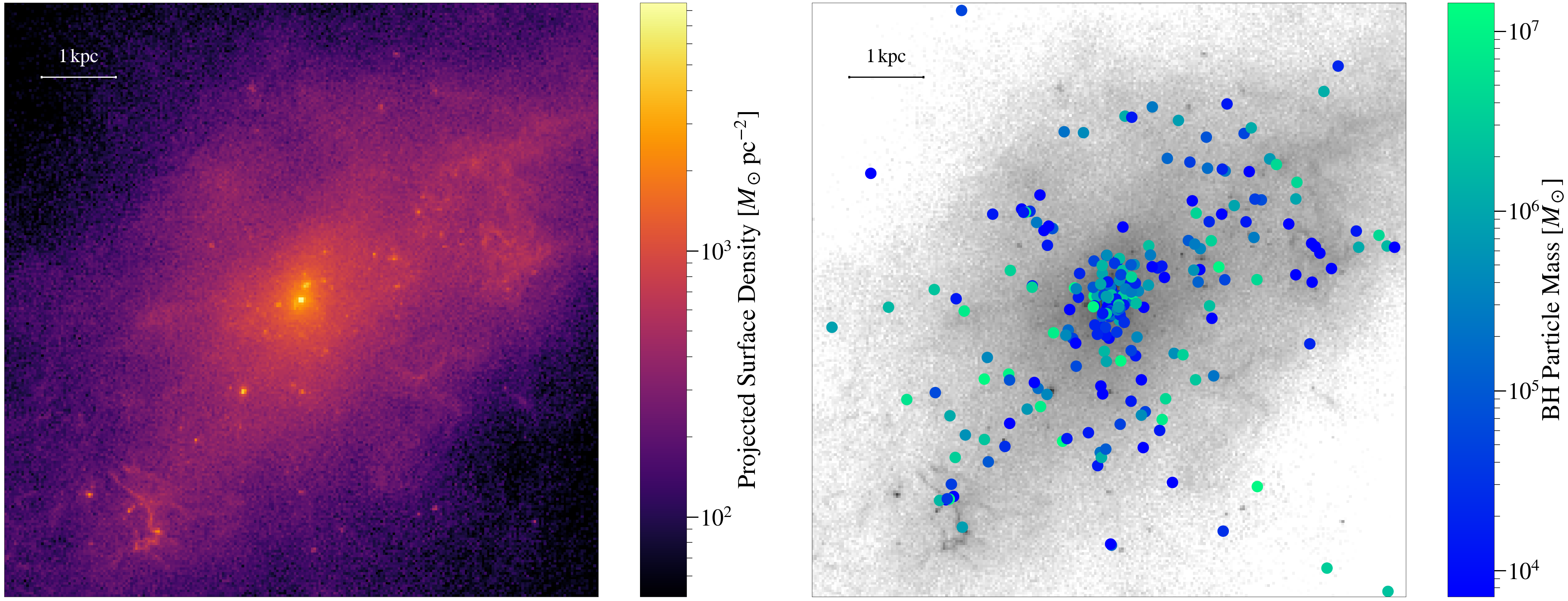}
    \caption{\textbf{Left:} Projected total non-BH mass (including dark matter, gas, and stars) density distribution of one of our simulations (``z9m12a'') at redshift $z=10.4$, as a typical simulation snapshot we analyze. The image shows the clumpy structure of high redshift galaxies. \textbf{Right:} The BH particles in this simulation at this particular snapshot, ranging from $10^3-10^7\,M_\odot$, covering a wide range of possible masses from different seed BH formation scenarios. BHs appear mostly randomly distributed in the galaxy, but with enhanced clustering near the galactic center. However we do not see significant seed-BH mass dependence, and the apparent galactic-center clustering simply reflects the overall concentration of mass (the galaxy effective radius here is $\sim\,$kpc).}
    \label{fig:galaxy}
\end{figure*}

\subsubsection{Simulation Details}

The simulations we study are re-simulations of the high-redshift ($z>5$) galaxies presented in \cite{xiangchengma2018,xiangchengma2017,xiangchengma2019} based on the Feedback In Realistic Environments
(FIRE; \citealt{hopkins:2013.fire,fire2}) project\footnote{See the {\small FIRE} project website: \FIREurl}. Specifically, we re-simulate the cosmological zoom-in simulations centered around the galaxies ``z9m12a'' and ``z5m12b''. Each of these represents a galaxy which has reached a halo mass $\gtrsim 10^{12}\,M_{\odot}$, a stellar mass $> 10^{10}\,M_{\odot}$, and a star formation rate $\gtrsim 150\,M_{\odot}\,{\rm yr^{-1}}$ by redshifts $z\sim 9$ and $5$, respectively. As discussed in \citet{xiangchengma2019}, these are chosen to be plausible analogues to the observed hosts of the highest-redshift, brightest QSOs. We note that while there are many other well-resolved galaxies in each cosmological zoom-in volume, we follow the most massive galaxy as it is the best candidate for a QSO host (but our conclusions about failure of BHs to ``sink'' are even stronger in lower-mass galaxies).

The simulations are run with an identical version of the {\small GIZMO}\footnote{A public version of {\small GIZMO} is available at \gizmourl} code \citep{gizmo} to their original versions in \citet{xiangchengma2017}. We use the mesh-less finite-mass (MFM) mode for solving hydrodynamic equations, with the identical FIRE-2 implementation of star formation and stellar feedback. The detailed baryonic physics included are all described extensively in \citet{fire2}, but briefly summarized here. Gas cooling includes a variety of processes (molecular, atomic, fine structure, recombination, dust, free-free, Compton, etc.) accounting for 11 separately tracked species (H, He, C, N, O, Ne, Mg, Si, S, Ca, and Fe), following the meta-galactic UV background from \cite{faucher-giguere2009} with self-shielding. Stars are formed on the free-fall time from gas which is locally self-gravitating, molecular/self-shielded, denser than $n > 1000\,{\rm cm^{-3}}$, and Jeans-unstable following \citet{hopkins2013}. Each star particle, once-formed, represents an IMF-sampled population of known mass, age and metallicity, and we explicitly account for stellar mass-loss (from OB and AGB outflows), core-collapse and Ia supernovae, and radiative feedback (in the forms of photo-ionization and photo-electric heating, and single and multiple-scattering radiation pressure), with rates tabulated from standard stellar evolution models \citep{Leitherer1999}.

The only difference between our simulations and those in \cite{xiangchengma2017} is that we re-run them including a ``live'' model for the formation of a broad spectrum of BH seeds, which are allowed to follow the full $N$-body dynamics. We emphasize that we do not artificially ``force'' BHs to follow the potential minimum or decay their orbits via any prescriptions of sub-grid DF, as in some cosmological simulations (e.g. \citealt{Springel2005bh,Hopkins2005,Hopkins2006,Hopkins2008,sijacki2015,Angles-Alcazar2017}).

We form BH seeds as follows: whenever gas meets all the star formation criteria above and is about to be transformed into a star particle, it is assigned a probability of instead becoming a BH seed. Instead of setting the probability as an adjustable constant as in, e.g. \cite{Bellovary2011}, it is weighted so that BH seeds form preferentially at the lowest metallicities \citep{Tremmel2017} and highest surface densities/gravitational acceleration scales: specifically, we adopt $p \propto \exp{(-Z/0.01\,Z_{\odot})}\,[1-\exp{(-\Sigma/\Sigma_{0})}]$ where $\Sigma \sim M/R^{2}$ is integrated to infinity with the Sobolev estimator from \citet{fire2} and $\Sigma_{0}=1\,{\rm g\,cm^{-2}}$, with $0.01\,Z_{\odot} = 1.4\times10^{-4}$. The metallicity weighting is motivated to be consistent with our current understanding of seed BH formation models, all requiring low-metallicities. For instance, Pop III stars and direct collapse models require low-metallicity primordial gas, while models of runaway mergers in star clusters strongly favor low-metallicity due to the lower mass-loss of massive stars in such environments \citep{Gonzalez2020}. The value of $\Sigma_0$ is specifically chosen because it is the density where analytic models \citep{Fall2010} and numerical simulations \citep{Geen2017,grudic:sfe.cluster.form.surface.density,kim2019} of individual star formation and BH growth have shown robustly that stellar feedback fails to ``blow out'' gas from the region efficiently, leading to runaway collapse/accretion. Exceeding this limit is required in many (but not all) models for massive BH seeds, either to prevent extended accretion disks from being destroyed by radiation from the accreting proto-quasi-star in direct collapse models, or as a necessary requirement to form super-dense star clusters, which are the essential prerequisite for star cluster-based IMBH formation models (e.g. runaway merging) to initiate rapid growth (see e.g. \citealt{shi2020,grudic:sfe.cluster.form.surface.density}).
The normalization of $p$ is chosen to form the maximum number of seeds before they begin to represent an appreciable fraction of the total galaxy mass and therefore perturb the dynamics. If the particle is selected to become a BH seed, then we draw a BH seed mass uniformly in $\log{M}$ from $M=10^{3}-10^{7}\,M_{\odot}$.

Because we wish to {\em only} study the dynamics of BH seeds, we ignore BH accretion or feedback. These will be studied in future work.

\subsubsection{Resolution and Treatments of (Un)Resolved DF}

Our ``default'' simulations have an approximately constant baryonic mass resolution of $\Delta m_{i}\sim 7000\,M_{\odot}$ and a 5 times higher DM resolution. This is sufficient to explicitly resolve N-body DF and other effects on the more massive seeds ($\gtrsim 10^{5}\,M_{\odot}$) we simulate: depending on the details of the gravity scheme, one generally achieves this for seed masses $M \gtrsim (10-100)\,\Delta m_{i}$.\footnote{We enable the additional improvements to the gravitational timestep criteria, tidal force treatment, tree-opening, and integration accuracy detailed in \citet{Guszejnov2020,2020MNRAS.495.4306G} where they were developed for simulations of star formation which require accurate evolution of stellar binaries and multiples, and set the force softening of the BH seeds to a very small value ($10^{-3}$\,pc) to represent real sink particles while using adaptive force softening for all other types to represent a smooth background. Detailed studies have shown that using adaptive softening as we do to ensure a smooth background force and with the more strict timestep and integration accuracy criteria used here, DF-like forces can be accurately captured for BHs with masses $\gtrsim10$ times the background particle mass, while with less accurate integration often used in cosmological simulations which do not intend to resolve few-body effects, the pre-factor is more like $\sim 100$ \citep{vandenbosch:no.orbital.circularization.due.to.dyn.frict,colpi:2007.binary.in.mgrs,boylankolchin:merger.time.calibration,fire2,2019MNRAS.486..101P,Barausse2020,2020MNRAS.495L..12B}.}
To assess the effects of resolution on the dynamics of lower-mass BH seeds, we briefly re-simulate one of our galaxies after applying a super-Lagrangian (AMR-like) refinement step (e.g. \citealt{Angles-Alcazar2020}), to run with $800\,M_{\odot}$ baryonic resolution
\footnote{Since the gravitational acceleration for BHs we study is strongly dominated by baryonic masses near the galactic center (we confirm the N-body forces from dark matter are sub-dominant by order-of-magnitude or more), we did not refine the dark matter mass/force resolution in these re-simulations, as it is largely irrelevant to our conclusions.}
, and measure whether there is any significant difference in the ``sinking rate'' of seeds at any BH mass after $100\,$Myr. We find no measurable difference. There is a simple reason why the detailed numerical accuracy of the DF forces on such low mass seeds has little effect: the actual DF time for low-mass seeds (with e.g. $M \ll 10^{5}\,M_{\odot}$) is far longer than the Hubble time at these (high) redshifts, so DF plays an essentially negligible role in their dynamics on a {\em galactic} scale.

\subsection{Semi-Analytic Orbital Evolution}
\label{sec:methods:analytical}

Several authors who have implemented DF as a sub-grid routine (e.g., \citealt{2019MNRAS.486..101P}) pointed out that sub-grid corrections could make a difference in the seed BH orbits. This may also be an issue for the accuracy of direct simulations, especially for low-mass seed dynamics. It is therefore useful to check the validity of our simulations with some alternative approach. Hence, we implement a semi-analytic analysis for the dynamics of BH seeds in post-processing, both as a check of our direct numerical simulations, and a way to gain analytic insight and explore even larger parameter spaces prohibited by the resolution and computational expense of our simulations. In post-processing, we can create an arbitrary sample of BH seeds at any desired time, and evolve them in time-independent potentials taken directly from the numerical simulations, allowing us to map the dynamics in detail.

To do so, we re-calculate the trajectories of 100 BH ``test particles,'' taking background potentials from the simulations and adding an analytic DF force explicitly in post-processing, during which we apply a newly developed DF estimator that is discussed in a companion paper (Ma et al. in prep.). We approximate the N-body dynamics of a seed of mass $M$ with an acceleration ${\bf a}_{M} = {\bf a}_{\rm ext} + {\bf a}_{\rm df}$, where ${\bf a}_{\rm ext}$ is the ``normal'' external gravitational acceleration on a test particle (computed identically to how the forces are computed in-code, for the adaptively force-softened potential from all N-body particles in the simulation). Then ${\bf a}_{\rm df}$ is the ``DF force'' -- the next-order (non-linear) term which represents the drag force arising from deflection of bodies by $M$. Specifically, we adopt the following expressions which can be {\em directly} computed from the simulation data (either on the fly or in post-processing):
\beq
\label{eqn:accel.full}
\begin{split}
{\bf a}_{\rm ext} =& -\sum_{i} \left( S_{i}(r_{i})\,\frac{G\,\Delta m_{i}}{r_{i}^{2}} \right)\,\hat{\bf r}_{i} \\ 
{\bf a}_{\rm df} =& \sum_{i} \left( \frac{\alpha_{i}\,b_{i}}{(1+\alpha_{i}^{2})\,r_{i}} \right)\,\left( S_{i}(r_{i})\,\frac{G\,\Delta m_{i}}{r_{i}^{2}} \right)\,\hat{\bf V}_{i}
\end{split}
\eeq
Here ${\bf a}_{\rm ext}$ and ${\bf a}_{\rm df}$ are defined as a sum over all N-body particles $i$, with N-body masses $\Delta m_{i}$, relative position ${\bf r}_{i} \equiv {\bf x}_{i} - {\bf x}_{M}$, relative velocity ${\bf V}_{i} \equiv {\bf v}_{i} - {\bf v}_{M}$, $v \equiv |{\bf v}|$ and $\hat{\bf v}={\bf v}/v$, with $G$ the gravitational constant, $\alpha_{i} \equiv b_{i}\,V_{i}^{2}/G\,M$ dimensionlessly parameterizing encounter strength, and $b_{i} \equiv r_{i}\,|\hat{\bf r}_{i} - (\hat{\bf r}_{i}\cdot \hat{\bf V}_{i})\,\hat{\bf V}_{i}|$ the impact parameter. $S_{i}(r_{i})$ is the usual dimensionless force-softening kernel to prevent numerical divergences, defined as
\beq
\label{eqn:force.softening}S_{i}(r_i)= \begin{cases}
\frac{32}{3}q_i^3-\frac{192}{5}q_i^5+32q_i^6\qquad& 0\leq q_i<\frac12 \\
-\frac{1}{15}+\frac{64}{3}q_i^3-48q_i^4+\frac{192}{5}q_i^5-\frac{32}{3}q_i^6\qquad& \frac12\leq q_i<1 \\
1\qquad& q_i\geq1 \\
\end{cases}
\eeq

We refer interested readers in our expression for ${\bf a}_{\rm df}$ to the companion paper (Ma et al. in prep.). But briefly, our expression reproduces exactly the classical \citet{chandrasekhar1943} (C43) expression ${\bf a}_{\rm df}^{\rm C43} = 4\pi\,G^{2}\,M\,\rho\,\ln{(\Lambda)}\,V^{-2}\,[{\rm erf}(V/\sqrt{2}\sigma) - (2/\pi)^{1/2}\,(V/\sigma)\,{\rm exp}(-V^{2}/2\,\sigma^{2})]\,\hat{\bf V}$ in cases consistent with the assumptions of C43, i.e.\ when the background distribution function is spatially homogeneous (constant density and velocity), time-invariant, Maxwellian, and single-component. But it allows more naturally for cases which violate these conditions. Our expression also removes the ambiguity of the C43 expression in estimating a number of ill-defined continuum quantities, when applied to discrete simulation N-body data (e.g.\ how and on what scales to evaluate $\rho$, $\sigma$, $V$; what value of $\Lambda$ to use). Usually, $\alpha_i\gg1$ such that ${\bf a}_{\rm df}\propto\sum\alpha_i^{-1}\propto M$, which means as expected that the DF acceleration is the largest for the most massive BHs, and potentially negligible for small BHs.

\section{Results}
\label{sec:results}

\subsection{Direct Simulations}
\label{sec:results:sims}

\begin{figure}
	\includegraphics[width=\columnwidth]{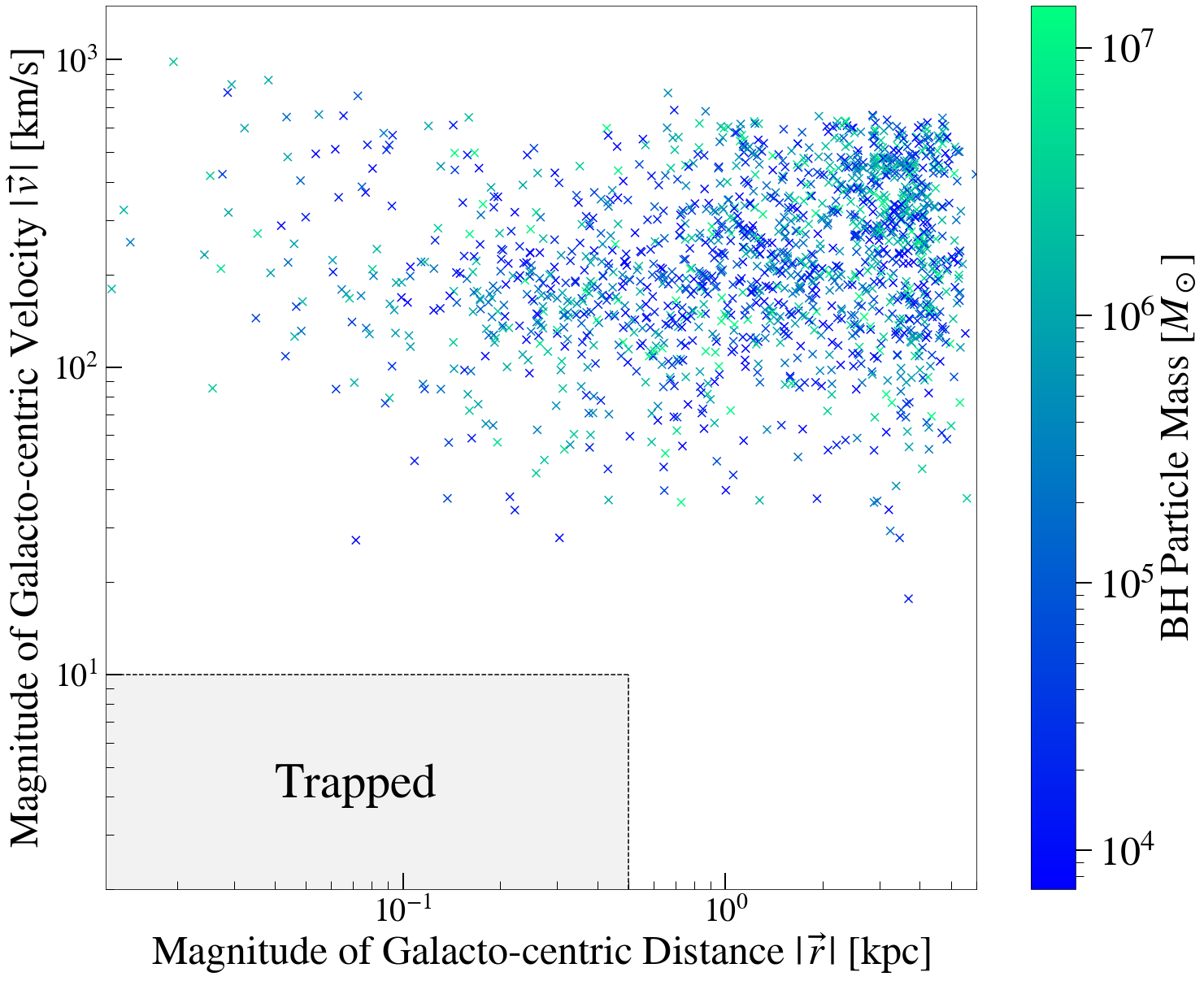}
    \caption{The magnitudes of velocities and galacto-centric distances of simulated BH particles for a general selection of snapshots in our simulations. We define a BH particle being trapped and efficiently sinking if it is located within $<0.5$\,kpc from the galactic center with a speed less than $10\,{\rm km\,s^{-1}}$ (shaded area). The colors label the mass of each BH particle. From our simulations there are no BH particles trapped in this manner, nor any significant dependence on their masses of their positions and velocities.}
    \label{fig:bh_phase_space}
\end{figure}

Here we present the results from direct simulations, focusing on the clustering behaviour of BH particles. In Figure \ref{fig:galaxy} we show a projected image of the galaxy ``z9m12a'' at redshift $z=10.4$, as a typical high redshift snapshot in our simulations. The left panel shows the total non-BH mass (i.e., including dark matter, gas, and stars) density distribution, with the galactic center located at the origin. The image shows the extremely clumpy appearance of typical high-$z$ galaxies, with multiple local density maxima near the galactic center, consistent with both other simulations and observations. In the right panel, we over-plot the positions of BH particles near the galactic center. The color labels their masses, ranging from $10^3-10^7\,M_\odot$, which cover a wide range of seed BH masses from different formation scenarios. There is no significant position dependence upon mass for BH particles in the galaxy, with some mild clustering near the galactic center. No significant mass dependence is observed.

To analyse the sinking problem of seed BHs, we show the magnitudes of galacto-centric distance $\mathbf{r}$ and velocity $\mathbf{v}$ of BH particles selected from 9 different snapshots in Figure \ref{fig:bh_phase_space}. Specifically, the BH particles are selected from snapshots in ``z5m12b'' at $z=9.0, 7.7, 7.0, 5.9$ and $5.0$, and snapshots from ``z9m12a'' at $z=10.9, 10.4, 9.9$ and $9.5$. Although snapshots at later redshifts contain BH particles that are already present at earlier redshifts in the same galaxy, the different snapshots are well separated in time such that the positions and velocities of these BH particles can be considered to be statistically independent. If a BH particle is located within 0.5 kpc from the galactic center with a (relative) velocity less than 10 km/s (Figure \ref{fig:bh_phase_space} shaded area), we consider it to have ``efficiently'' undergone sinking and trapping in the galactic center. Figure \ref{fig:bh_phase_space} suggests that none of our BH particles in the mass range of $10^3-10^7\,M_\odot$ has achieved this at the redshift they are observed. There is also no clear dependence of BH positions and velocities on their masses, indicating their dynamics is basically independent of their masses if BH masses are below $10^7\,M_\odot$, i.e. the dynamics is dominated by the mass-independent external gravity, while the mass-dependent DF plays a negligible role.

\subsection{Semi-Analytic Orbital Evolution}
\label{sec:results:analytic}

\begin{figure*}
    \centering
    \includegraphics[width=\textwidth]{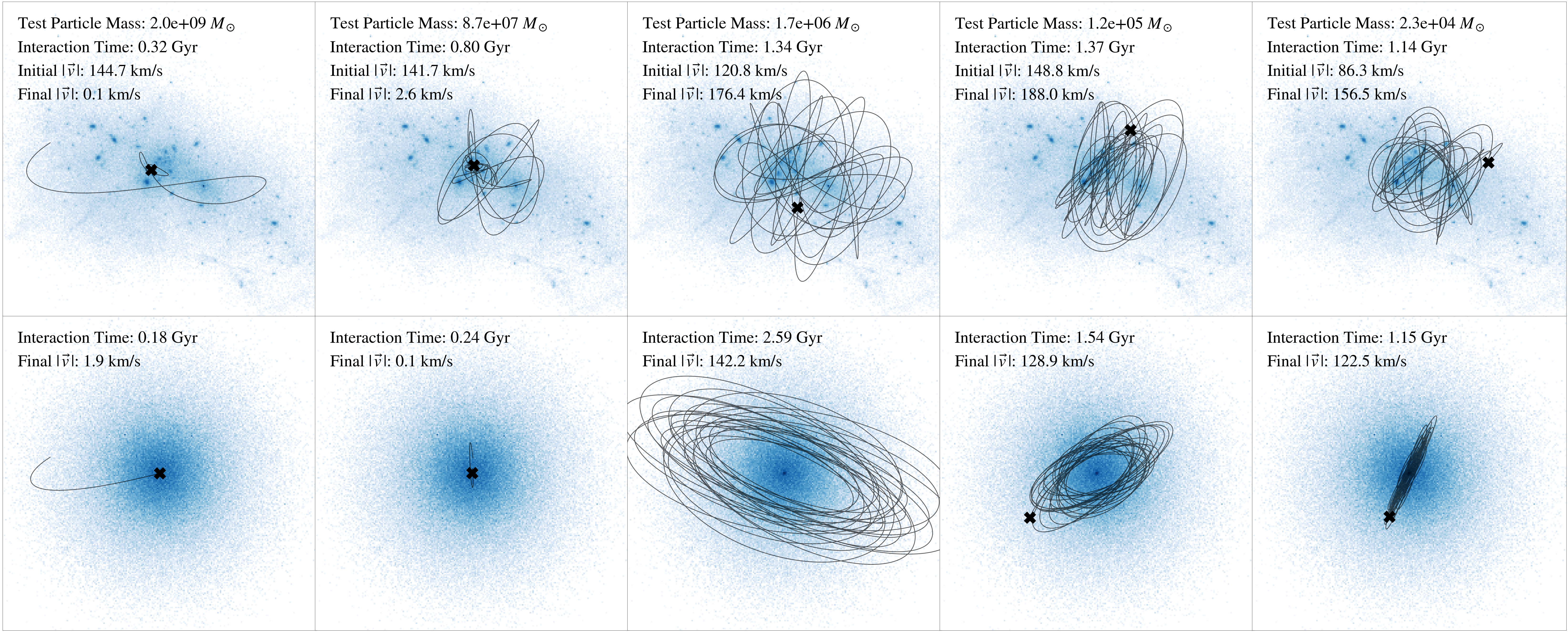}
    \caption{Sample orbits of several test particles (overlaid on top of the mass density distribution, shown in the blue colorscale) in the $z=7$ snapshot of ``z5m12b'' ({\bf upper}) and its ``spherically smoothed equivalent'' ({\bf lower}) where we take the identical enclosed mass profile $M_{\rm enc}(<r)$ and re-distribute the mass to be a perfectly-spherically-symmetric potential. The thin lines show the trajectories and the black cross shows the final positions of test particles. Each panel is $8$ kpc across in spatial scale. We find that in the high-$z$ galaxy, the most massive test particles do sink to the galactic center within a Hubble time at $z=7$ ($\sim1\,\text{Gyr}$), while the low-mass seeds are simply experiencing chaotic orbits. In the smooth galaxy, the sinking behaviour is not very different for these five samples, yet for the massive seeds which are able to sink, their sinking time reduces drastically. This suggests that clumpy galactic backgrounds generally inhibit the sinking of massive seeds.}
    \label{fig:orbits}
\end{figure*}

\begin{figure}
	\includegraphics[width=\columnwidth]{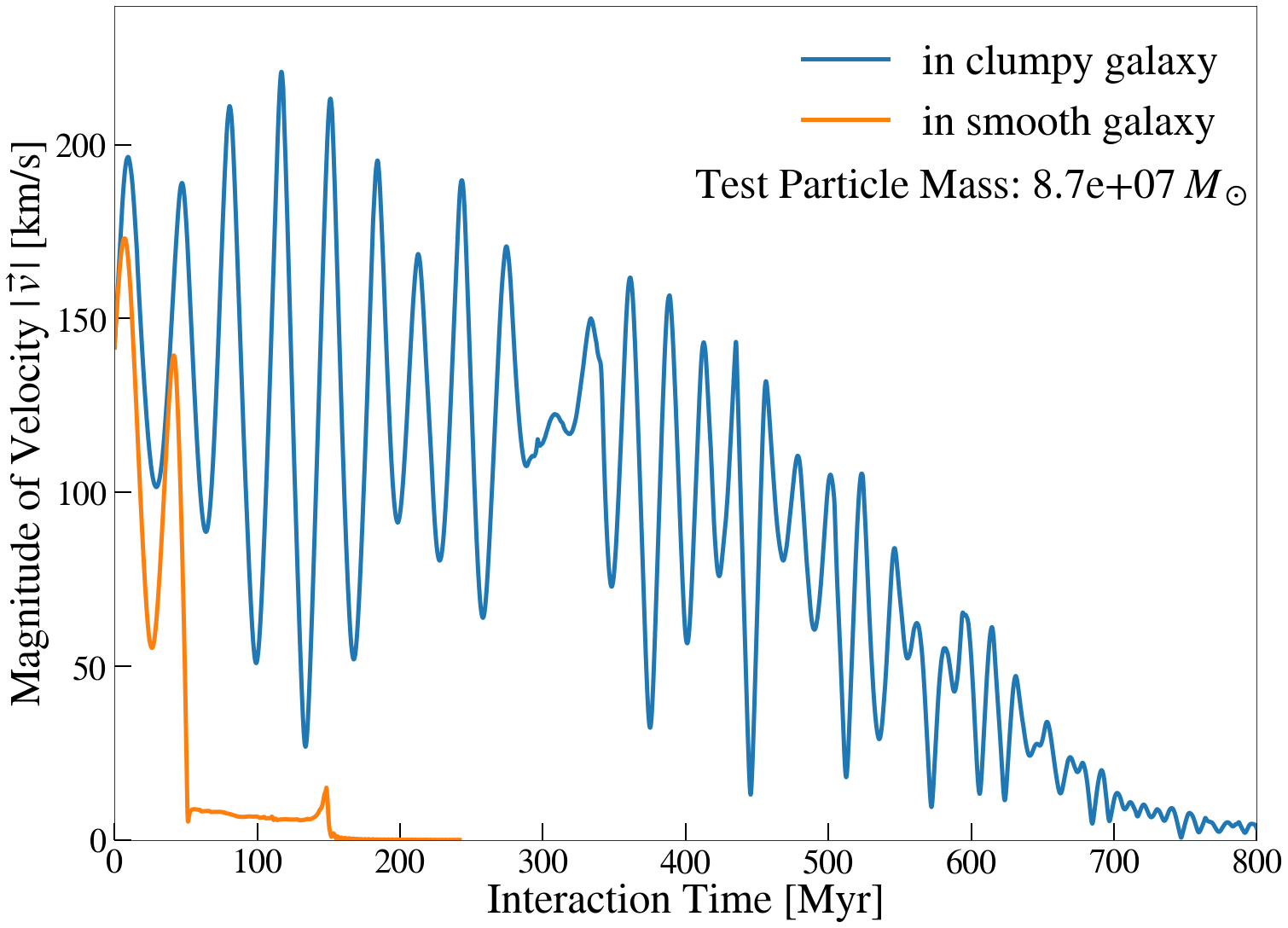}
    \caption{The evolution of the magnitude of the BH velocity as a function of interaction time for our integration of a $8.7\times10^7\,M_\odot$ test particle. We see that in both the clumpy and spherical smoothed galaxy, the velocity decays within 1 Gyr. But in the smooth galaxy the decay time is lower by about one order of magnitude than the clumpy case, suggesting again the clumpy and chaotic nature of early galaxies may drastically increase the sinking time of seed BHs.}
    \label{fig:test_particle_smooth_compare}
\end{figure}

\begin{figure*}
    \centering
    \includegraphics[width=\textwidth]{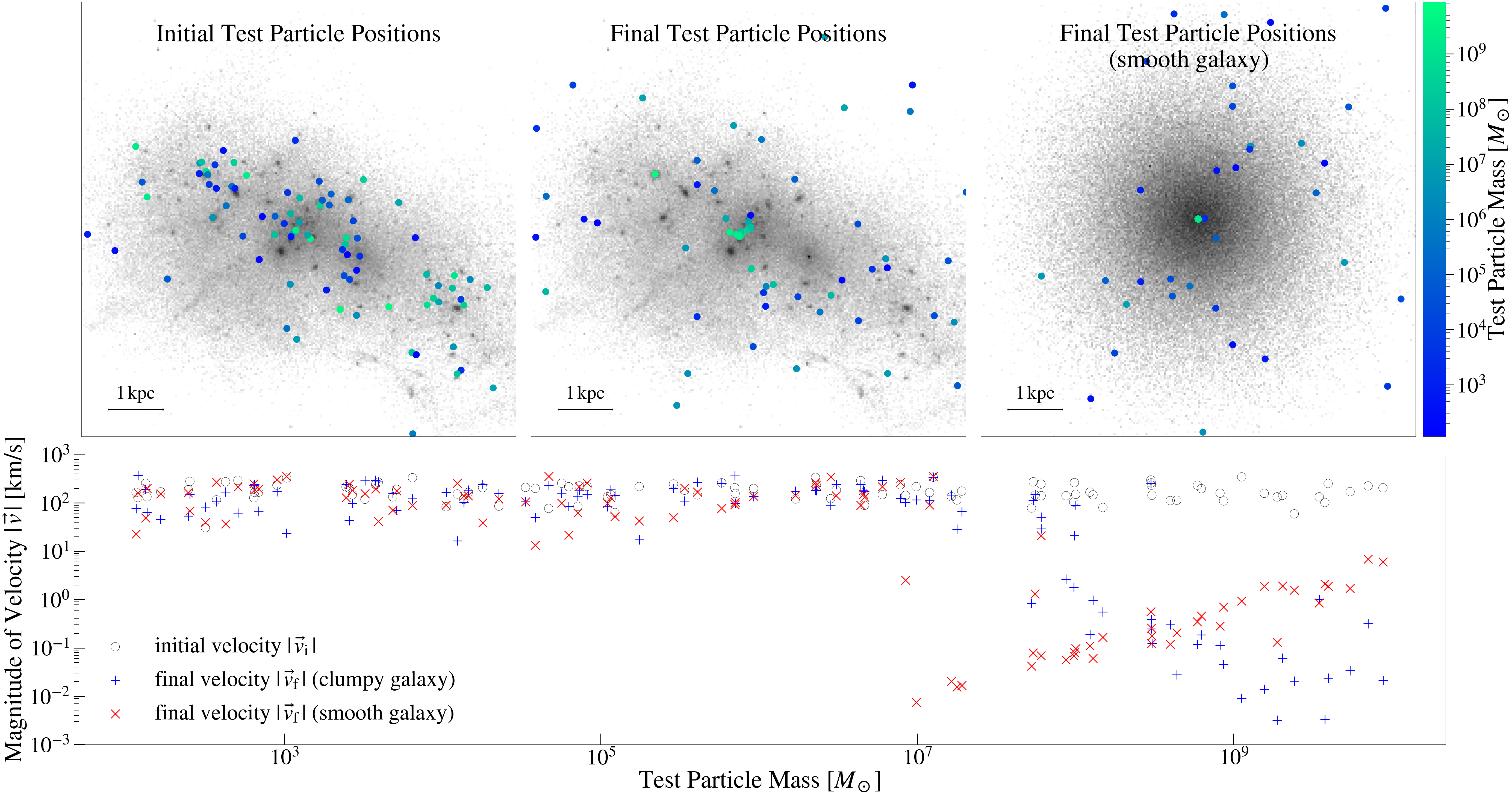}
    \caption{\textbf{Upper left:} The initial positions of test particles which we semi-analytically integrate, overlaid on the mass density distribution (grey) for ``z5m12b'' at redshift $z=7$; \textbf{Upper middle:} The final positions of these test particles. \textbf{Upper right:} The final positions of these test particles integrated in the spherically smoothed galaxy. The colors label the test particle masses. {\bf Lower:} The magnitude of initial velocities and final velocities as a function of the BH mass. We see that for the clumpy galaxy, the high mass ($M\gtrsim10^8\,M_\odot$) test particles sink to the galactic center after the integration, while the low mass particles remain randomly distributed. For the smooth galaxy the minimum mass for sinking reduces to $M\gtrsim10^7\,M_\odot$, about one order of magnitude lower. DF and sinking are negligible for the lower-mass seeds in both cases $^5$.}
    \label{fig:all_test_particles}
\end{figure*}

\begin{figure}
	\includegraphics[width=\columnwidth]{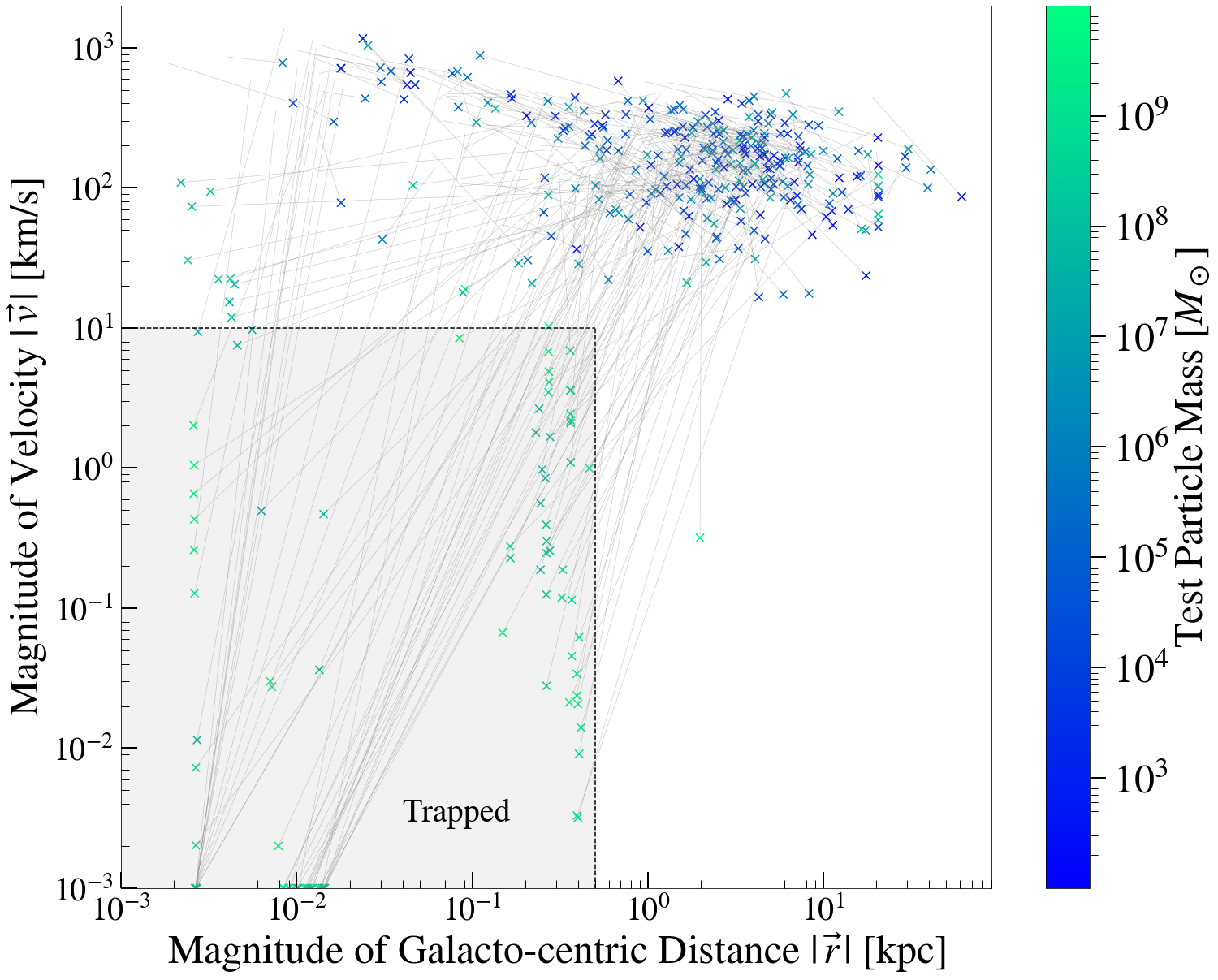}
    \caption{The initial and final magnitudes of velocities and galacto-centric distances of all our test particles across different snapshots. The colored points show the final velocities and distances (with any final velocities less than $10^{-3}$ km/s interpreted as $10^{-3}$ km/s for clarity). We define a BH particle as ``trapped'' as in Fig.~\ref{fig:bh_phase_space}. The thin grey line connects the final properties with initial properties of each particle. The colors label the mass of each particle. We can see that after our integration nearly all particles with masses $\gtrsim 10^8M_\odot$ sink to the galactic center (with a significant decline of velocity and distance), yet lower-mass particles are still randomly distributed.}
    \label{fig:test_particle_phase_space}
\end{figure}

Here we present the results from semi-analytic post processing, with our new DF estimator, to cover a wider range of BH masses. Specifically, we select snapshots from ``z5m12b'' at $z=5.0,7.0,9.0$ and ``z9m12a'' at $z=9.5$. In each snapshot, we place 100 test particles to integrate their dynamics, whose initial parameters are generated in the following way: the masses are randomly selected from $100-10^{10}\,M_\odot$ (uniformly sampling log of mass), while the initial positions and velocities are chosen randomly from star particles in the corresponding snapshot, which is not only a convenient sampling method, but physically motivated since we would expect seed BHs are mostly born in similar locations to star clusters. With such sampling, are also able to study the effects of initial galacto-centric distances/velocities on sinking (so we stress that our conclusions are completely independent of how we perform this sampling). In post processing, we ignore the dynamics of background particles, i.e. we apply a time-independent gravity potential, as we would expect the static background to represent random sample of typical chaotic high-$z$ galaxies, not an accurate reflection of some certain galaxy. The assumption of a static but realistically clumpy mass distribution allows us to gain insight into the effects of spatial inhomogeneities in the gravitational potential expected in typical, chaotic high-z galaxies. However, the orbits that we calculate in this way are not necessarily fully realistic since they neglect the time dependence of the potential. We note, though, that time dependence of the potential seems unlikely to accelerate sinking relative to a static-potential calculation; if anything time-dependence of the potential could further contribute to keeping seeds away from the galactic center. The external gravity and DF are calculated by Equation \ref{eqn:accel.full}.
Essentially, the difference between our ``live'' dynamics simulations and these post-processing calculations allows us to see how the time-dependence of the potential alters (in aggregate) the dynamics of sinking BH seeds. 

To further see how the ``clumpiness'' of the potential alters the BH dynamics, we re-run our semi-analytic orbit integration in a ``spherically-smoothed'' version of the potential. In these calculations, we take the exact same spherically-averaged mass profile from the full simulation snapshot studied above, $\rho(r) \equiv d M_{\rm enc}(<r) / 4\pi\,r^{2}\,d r$ in narrow radial annuli $dr$, and then use this as the background potential for our orbit integration. So, by definition, this has the same spherically-averaged $M_{\rm enc}(<r)$ and circular velocity profile, but no substructure.

In Figure \ref{fig:orbits} we show several sample orbits for test particles of different masses in the $z=7.0$ snapshot of ``z5m12b'' overlaid on its mass density distribution. The orbits in the original snapshots are shown in the upper panel, while in the lower panel we show the trajectories integrated from the spherically smoothed version of this snapshot, with the same test-particle initial conditions. The thin lines show the trajectories and the black cross shows the final positions of test particles. The test particles follow chaotic orbits in the clumpy snapshot with no significant dynamical center (as we would expect for a high-$z$ galaxy). It appears that for the most massive test particles $M\gtrsim10^8\,M_\odot$, their velocities significantly decrease within a Hubble time at $z=7$ ($\sim1\,\text{Gyr}$), and their final positions lie within the very central region of the galaxy. But there is no significant sinking for low-mass test particles. In the smooth galaxy the particles behave similarly, yet it takes a shorter interaction time for the most massive test particles to sink. The velocity evolution of one particular test particle of $8.7\times10^7 M_\odot$ is shown in Figure \ref{fig:test_particle_smooth_compare}, and it is shown that the velocity decay timescale is about one order of magnitude shorter in the smooth galaxy compared to the clumpy galaxy. This suggests that the clumpy nature of early galaxies may increase the sinking time of seed BHs by an order of magnitude, by introducing chaotic dynamics to their orbits. In Figure \ref{fig:all_test_particles} we show the initial and final positions of all test particles we integrate in this particular snapshot, and its spherically smoothed version. We also show their initial and final velocity magnitudes as a function of mass in the lower panel. In the clumpy galaxy, while the test particles are randomly distributed in the galaxy initially, those with $M\gtrsim10^8\,M_\odot$ show clustering behaviour near the center after the integration, and their speeds decay to less than a few kilometers per second, indicating that they sink to the galactic center after the integration. The remaining low-mass particles remain scattered around, with no significant decay of their speeds. The smooth potential reduces the minimum sinking mass to $\sim10^7\,M_\odot$, when test particles are integrated over an order of the Hubble time at $z=7$ \footnote{
There is a trend of increasing final speed with test particle mass in Figure \ref{fig:all_test_particles} for the smooth galaxy. This turns out to be a reflection of the different integration time of these particles: we apply a timestep control proportional to $|v/a|$ to avoid numerical errors. The massive particles, with larger DF (larger $a$), hence have smaller timesteps and shorter integration time compared to the less massive ones (see also the ``interaction time'' label in Figure \ref{fig:orbits}), experiencing less deceleration in the integration. This effect does not appear in the clumpy galaxy, since the lack of dynamical centers of these galaxies makes the particle dynamics chaotic, and the gravity and DF for these particles balance each other when they reach the center, making the interaction time less important.}. 

It appears that the clumpy nature of early galaxies may increase the ``minimum sinking mass'' by one order of magnitude. It is worth noting that the sinking massive particles in the clumpy galaxy also do not sink exactly to the same place near the center (as they do in the smooth galaxy). This implies that a clear definition of galactic center with resolution of a few hundred pc is still ambiguous for these galaxies, and has potentially major implications for the demographics of BH-BH mergers at high redshift.

In Figure \ref{fig:test_particle_phase_space} we show the initial and final magnitudes of galacto-centric distance $\mathbf{r}$ and velocity $\mathbf{v}$ of all our test particles across different snapshots. The colored points show the final velocities and distances of test particles while the thin grey lines connect their final values with initial values. We define the ``sinking'' region in phase space as in \S~\ref{sec:results:sims}. Since we are covering a larger mass range of test particles than what we did in direct simulations for BH particles, some of the most massive particles do efficiently sink to the ``trapped region'' this time. Specifically, particles with $M\gtrsim 10^8 M_\odot$ sink to the center region of the galaxy after the integration, regardless of their initial positions and velocities. For low-mass ($M \lesssim 10^8 M_\odot$) particles, their final position and velocity distributions appear to be statistically similar to their initial configurations. This confirms the robustness of our results from direct simulations, in which all BH particles are less than $10^8\,M_\odot$ and are therefore not experiencing significant sinking.

It is also worth noting that the sinking criterion almost depends {\em entirely} on the particle mass, not on initial velocities/distances to galactic center. This is in contrast to what one would naively infer from the simplest DF-time calculations which assume a smooth potential with a constant circular velocity and BHs on slowly-decaying nearly-circular orbits, in which case the sinking time depends explicitly on the initial distances $t_\mathrm{sink}\propto r^2$ \citep{BinneyTremaine}. Physically, this can be explained by three factors: (1) for highly-eccentric or radial orbits, the dependence on initial radius is much weaker, independent of the assumed density profile or details of the DF scaling \citep{hopkins:disk.heating}; (2) the chaotic dynamics of seed BHs in clumpy (i.e.\ non-smooth) galaxies effective erase the memories of their previous orbits, which makes the initial positions less important to their orbital decay; (3) the traditional $r^2$ dependence of $t_\mathrm{sink}$ depends explicitly on the implicitly-assumed isothermal mass density profile of the galaxy -- but more generally the DF acceleration scales as $a_\mathrm{DF}\propto\rho(r)/v_\mathrm{c}^2$. In a clumpy high-$z$ galaxy, however, the density $\rho$ is not necessarily falling rapidly as in an isothermal sphere (and is not a trivial smooth monotonic function of galacto-centric radius), again wiping out the naively-predicted $r$-dependence of $t_\mathrm{sink}$.

\section{Discussion}
\label{sec:discussion}

\subsection{Possible Solutions}

From both direct simulations and semi-analytic post-processing calculations, we have found that seed BHs less massive than $10^8\,M_\odot$ generally cannot sink to galactic centers via DF in high-$z$ galaxies. To have at least one seed BH positioned in the galactic center so that it could accrete to $\sim10^9\,M_\odot$ and provide a plausible origin for luminous high-redshift quasars, we discuss two categories of possible solutions.

\subsubsection*{Solution 1: A Large Number of Seeds, Forming Continuously}
\label{sec:discussion:lots.of.seeds}

\begin{figure*}
    \centering
	\includegraphics[width=\textwidth]{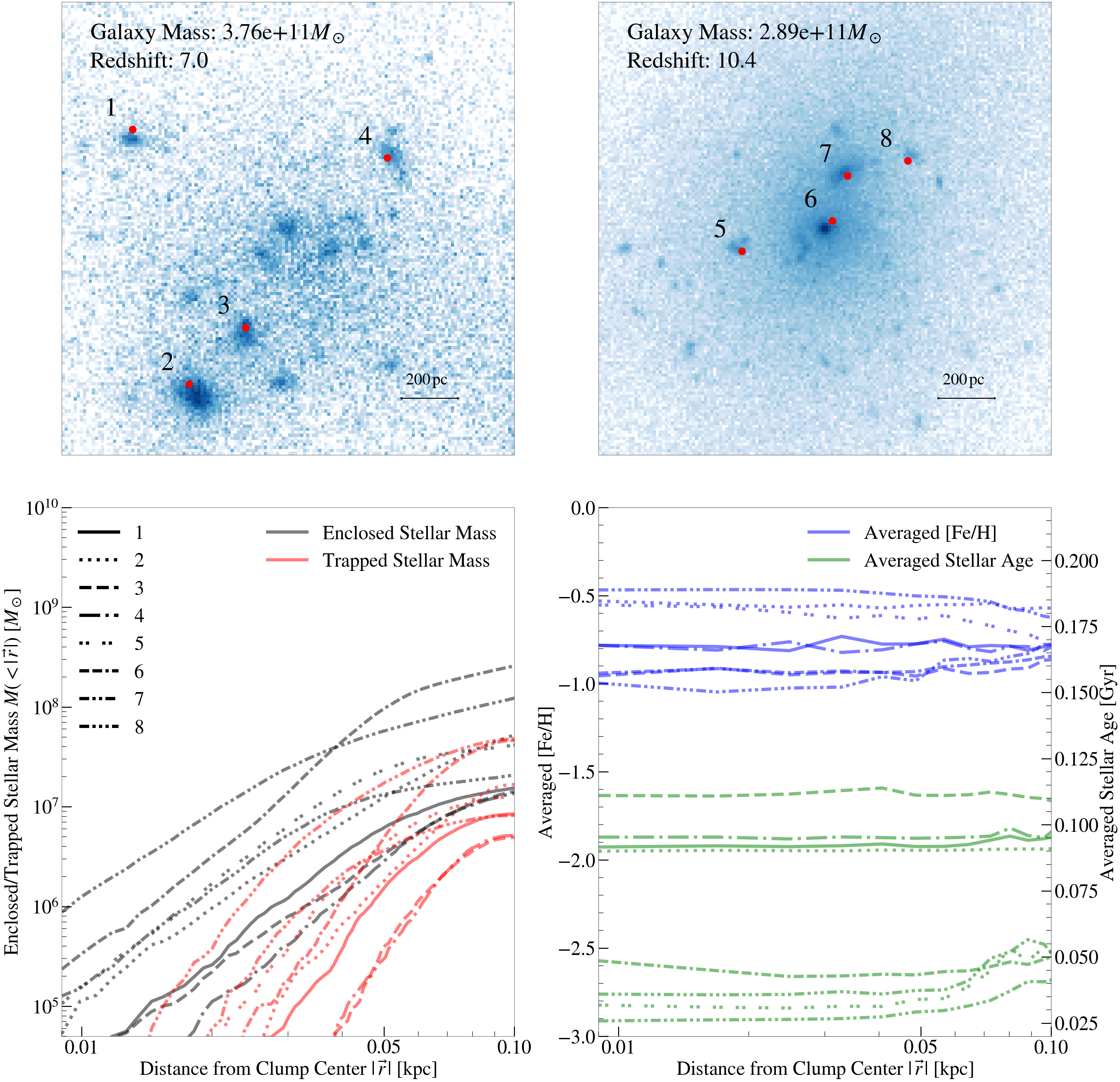}
    \caption{Behavior of low-mass test particles (e.g. stars) in individual high-density clumps (the ``proto-bulge'') within our simulations. 
    {\bf Upper Left:} Mass density distribution of a $z=7$ galaxy with a total matter mass of $3.8\times10^{11}\,M_\odot$, where clumps 1-4 (the most massive bound sub-structures) are identified. 
    {\bf Upper Right:} Same for a $z=10.4$ galaxy with a total matter mass of $2.9\times10^{11}\,M_\odot$, where clumps 5-8 are identified. 
    {\bf Lower Left:} Enclosed stellar mass inside each clump as a function of clump-centric distance, and the ``trapped'' mass (defined as the mass which is bound with apocentric radii inside this radius, as opposed to e.g. stars on ``plunging'' or unbound orbits; see text). 
    {\bf Lower Right:} Mean stellar metallicity and age for star particles inside each clump. 
    We see that only a few percent of the enclosed stellar particles could be trapped well inside ($|\mathbf{r}|\lesssim 50\,\mathrm{pc}$) the clumps. The metallicity and age also indicate that most star particles (hence the clump) are formed recently, which leads to new problems for some scenarios for seed BH growth.}
    \label{fig:clump_property}
\end{figure*}

The first option is to use numbers as a trade off for efficiency: although one low-mass seed BH is not likely to sink and accrete, a large number (which we estimate quantitatively below) of low-mass seeds could possibly give an opportunity for a ``lucky one'' to sink and grow. Since the dynamics of BH particles and star particles are identically solved in our simulations (both as collisionless dynamics with external gravity), and the masses of star particles are around $10^3 M_\odot$, below the low-mass end where DF drag is significant, we can use the star particles in our simulation as an ensemble of test particles to estimate the fraction of stars and therefore relics (ignoring processes like kicks) which can be trapped in local clustering structures (``clumps''). We apply such analysis to two particular snapshots, namely, ``z5m12b'' at $z=7.0$ and ``z9m12a'' at $z=10.4$.

We are only interested in clumps broadly near the galactic center, hence we identify the four densest clumps within 1.6 kpc near the galactic center for each snapshot respectively, as shown in the upper panels of Figure \ref{fig:clump_property}. The center of the clumps are identified as the local density maxima, and their geometrical shapes are treated as spherically-symmetric with radius $~100$ pc enclosing almost all of the clump mass, a fair approximation as shown in Figure \ref{fig:clump_property}.

The lower left panel of Figure \ref{fig:clump_property} shows the enclosed stellar mass and trapped stellar mass as a function of radius around each clump. If a star particle at radius $r$ has a maximum possible apocentric radius $r_\text{max}$ from the clump center (using the energy and angular momentum of each to evaluate its orbit, assuming the clump is static over its orbital timescale), we then say it is instantaneously enclosed within $r$ and ``trapped'' within $r_\text{max}$. The gravity potential is calculated assuming a static potential around each clump with spherical symmetry (the clumps themselves, by definition, do not have substantial substructure). We see that the stellar masses in each clump ($M_\mathrm{enclosed}(|\mathbf{r}|<100\,\mathrm{pc})$) range from $10^7$ to $10^8\,M_\odot$. The mass fractions of trapped stars differ for different clumps and around 30\%-50\% of stellar mass could be trapped in a $\sim0.1$\,kpc radius of the clumps, yet this value decreases as we go deeper into the clump center, and the clumps could eventually trap only a few percent of enclosed star particles within $\sim 50$\,pc. For all clumps, $\gtrsim 90\%$ of their mass is in stars (as opposed to gas or dark matter).  

\begin{figure}
	\includegraphics[width=\columnwidth]{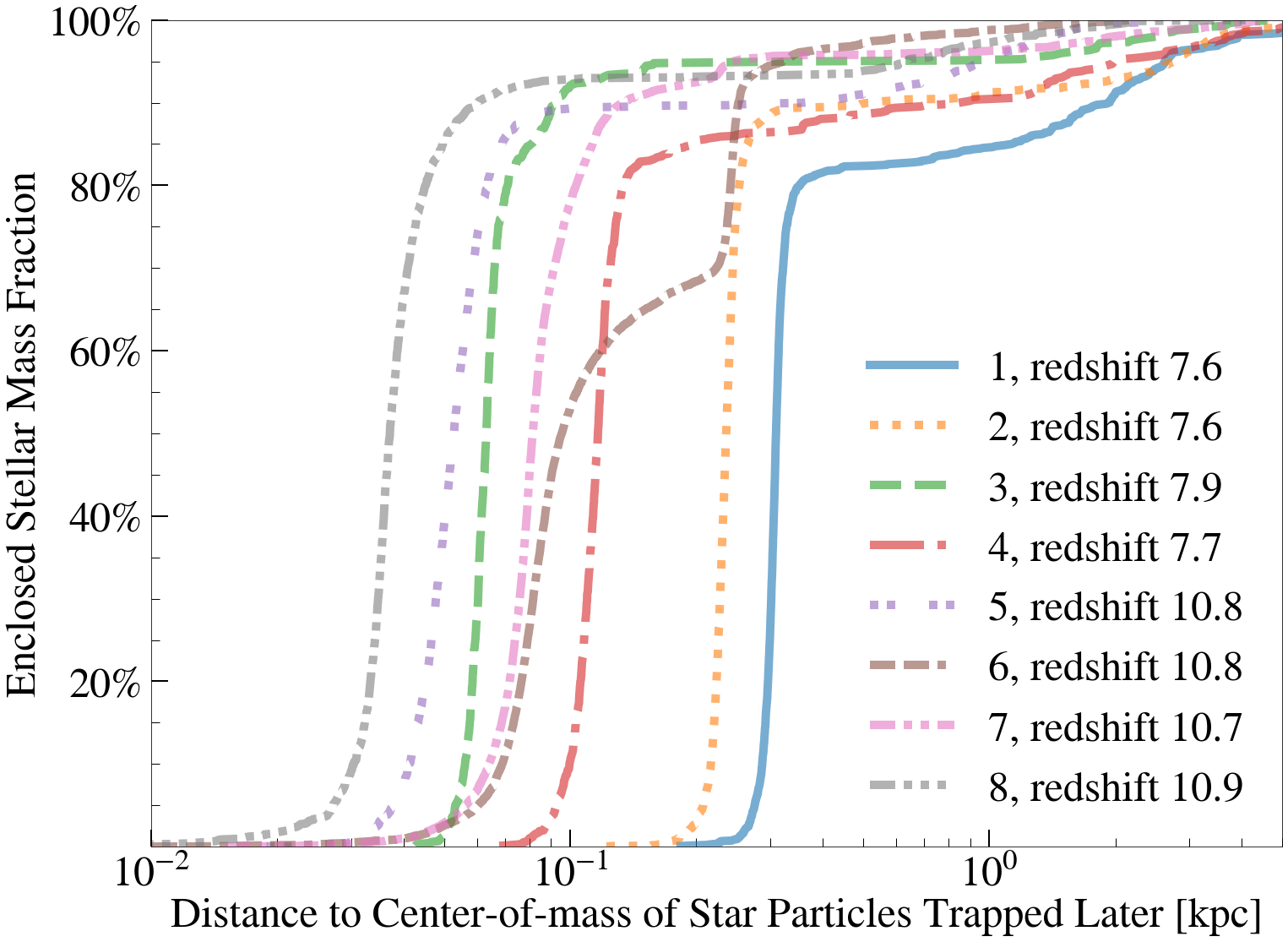}
    \caption{Distribution of ``formation distances'' for stars identified as enclosed in clumps in Fig.~\ref{fig:clump_property}. We plot the cumulative distribution of distances between the center-of-mass of the main clump progenitor and the newly-formed star particle, at the time each star particle formed. 
    We see that at least $>80-90\%$ of star particles in these clumps form ``in situ,'' at distances $\ll 1\,$kpc from the clump center. Only a small fraction are formed outside the clump and later captured. Of those, almost all form in the same galaxy at distances $<5\,$kpc (as opposed to in satellites or different progenitor galaxies).} \label{fig:previous_clumps}
\end{figure}

{\em Some} low-mass objects are trapped in the dense clumps that represent the proto-bulge of these galaxies. But do they actually ``sink'' or get trapped dynamically, or did they simply form in-situ? To track the formation history of these star particles, we show their distances to their center-of-mass at the particular redshift when most of them are just formed
\footnote{The simulations we use generate one snapshot per $~0.01$ scalefactor, which is sufficient for this exercise.}
in Figure \ref{fig:previous_clumps}. It turns out for almost all clumps, $>80-90\%$ of the star particles which we defined as ``trapped'' in these clumps are formed within $\ll 1\,$kpc from the clump-progenitor center-of-mass, which means most trapped star particles are formed in-situ. The only seemingly exception is clump 6, where at first glance it appears that only about $\sim 70\%$ of the trapped star particles are in situ particles, but a detailed analysis shows that the remaining particles are actually formed in another clump which merges with clump 6, which does not challenge the conclusion (though it does relate to the hypothesis discussed in \S~\ref{sec:discussion:giant.seeds}). Taken together, this means that while it is possible in principle for ``lucky'' low-mass objects to be ``trapped,'' it is quite rare: comparing the total stellar mass of the galaxy to the mass of stars which form ex-situ and are trapped near clump centers yields a probability of about $\sim 10^{-5} - 10^{-3}$ (depending on how generously we define ``trapped'') for a low-mass seed formed randomly in the galaxy to migrate to being ``trapped'' in the central $<100\,$pc of a clump by $z\sim 7$. 

Even if this occurs, the metallicity of the star particles which undergo this processes may create new problems for seed models. While the first Pop III stars or ``direct collapse primordial clouds,'' which are candidates for forming massive seed BHs, could form very early at metallicities $Z \ll 10^{-5}\,Z_\odot$, the metallicity of star particles enclosed/trapped in clumps (even restricting to the ``ex situ'' stars) is generally much higher, and turns out to be the highest for the most massive clump, as shown in the lower right panel of Figure \ref{fig:clump_property}. This indicates that the trapped star particles in these clumps may not represent a fair sample of the ex-situ seed BH particles which are formed before the clumps themselves are formed. The earliest-forming stars are actually the {\em least} likely to be trapped in such clumps: they tend to form in mini-halos at much earlier times and therefore across many different progenitors and thus have to migrate in from the furthest distances, while the ``ex situ but trapped'' stars primarily still form in situ ({\em in the same galaxy}) just at distances of $\sim 1\,$kpc from the clump.

For all seed BHs, either in-situ or ex-situ, a related problem is related to the tension between the required clump masses and their ages. In many SMBH formation mechanisms, seed BHs have a higher probability both to be initially trapped and to subsequently accrete gas rapidly in the most dense/massive clumps, but these clumps are preferentially formed later, hence providing less time for BHs to migrate and to accrete. The average age of star particles inside clumps, as shown in the lower right panel of Figure \ref{fig:clump_property}, is far less than the Hubble time at the redshift we examined, providing a strict constraint on duty cycle if seed BHs are indeed hyper-Eddington accreting to become SMBHs in these clumps. Nevertheless, it is worth noting that SMBH seeding prescriptions are still highly uncertain, and other mechanisms may be able to circumvent these constraints.

\subsubsection*{Solution 2: High ``Effective Masses'' for Seeds}
\label{sec:discussion:giant.seeds}

From the semi-analytic calculations in section \ref{sec:results:analytic} we have found that only seed particles as massive as $\gtrsim10^8\,M_\odot$ can efficiently or reliably sink to galactic centers in a Hubble time. Such a large mass, however, is already a SMBH. On the other hand, our analysis in the previous section has shown that dense young star clusters as massive as $10^7-10^{8}\,M_\odot$ {\em are} present near the galactic center. In the previous section we also show that most trapped star particles within those clumps are already formed in situ. This suggests another possibility: while randomly formed seed BHs are generally not massive enough to decelerate individually via DF, their preferential formation in tightly bound structures with large ``effective mass'' is more realistic, as clusters could scatter with other components in the galaxy and sink effectively to the galactic center. Indeed, in X.\ Ma et al., in prep.\ we show that the most-massive clumps do merge efficiently as these simulations are run to lower redshift and form the ``proto-bulge'' of the galaxy.

There have been numerous papers arguing that runaway mergers in dense globular (star) clusters are a potential way to produce intermediate mass black holes (IMBHs, with typical masses $10^2-10^5\,M_\odot$, see, e.g. \citealt{PortegiesZwart2002,Gurkan2004,shi2020,Gonzalez2020}), which naturally becomes a preferential way to embed massive BH seeds in dense clusters as described above. Such channels, however, suffer from other problems like large gravitational recoils that can remove the formed IMBHs from the cluster (e.g. \citealt{Holley-Bockelmann2008}). There are also works arguing that gas accretion in nuclear star clusters (NSCs) and starburst clusters can also build up the mass of IMBHs rapidly \citep{Kroupa2020,Natarajan2021}, which could be another way to apply this solution here. Yet observations have put upper limits on IMBHs masses (e.g. \citealt{Lutzgendorf2013,Lutzgendorf2015,Kamann2016,Zocchi2017}), which introduce additional constraints on these channels. It should also be noted that, while globular clusters are usually assumed to be mainly pristine clusters that formed at very high redshift in mini-halos, hence define an ``old'' population for astrophysicists in the local universe, they are not so much older than the stars at $z\gtrsim7$. In fact, the overwhelming majority of the clusters form {\em in-situ} in the galaxy as it evolves from in-situ gas, not from mini-halos merging in. This means that the metallicity and timing problems discussed in \S~\ref{sec:discussion:lots.of.seeds} apply to this scenario, as well.

\subsection{Comparisons to Other Works}

Our conclusions are consistent with other recent works focusing on slightly different aspects of this problem. For instance, \cite{Roskar2015} and \cite{Tamburello2017} study the co-evolution of SMBH pairs, finding that galactic clumps (originated either from high-$z$ star forming regions or a clumpy interstellar medium created by galaxy mergers) significantly perturb their orbital evolution, which potentially delay the decay process. \cite{Tamburello2017} and \cite{Tamfal2018} also point out that SMBH/IMBH pairs are still separated by $0.1-2\,\mathrm{kpc}$ after $\sim1\,\mathrm{Gyr}$ in their simulations, which is consistent with our findings that no well-defined galactic centers can be identified on sub-kpc scales under these conditions. \cite{Bortolas2020} simulate a $10^6\,M_\odot$ BH in a {\em non-clumpy} galaxy embedded in a cosmological environment at $z=6-7$ and they show that DF torques are usually unimportant compared to the large-scale stochastic gravitational torques in determining the BH decay, even if no clumpy structures are considered. These works support to our conclusion that the chaotic structures of high-$z$ galaxies could drastically change the sinking timescale (hence the minimum sinking mass), if only DF is considered.

\cite{2019MNRAS.486..101P} presented a complementary study to ours, focusing on more idealized simulations analogous to lower-redshift systems, and a smaller number of test cases, but considering in more detail many of the numerical details of ``live'' sub-grid BH DF treatments (e.g.\ explicitly adding an analytic DF force term in low-resolution simulations). They concluded that {\em even in} idealized galaxies designed by construction with a well-defined dynamical center and a single, massive, centrally-peaked bulge (e.g., an exponential-disk and an Hernquist bulge), lower-level clumpiness in the gas (e.g., GMCs with typical masses $\sim 10^{5}-10^{6}\,M_{\odot}$) would drive wandering or ejection of BHs with seeds less massive than $\sim 10^{5}\,M_{\odot}$. They hence concluded that $10^{5}\,M_{\odot}$ is the minimum required mass for a BH to be well stabilized in the center of its host. Since observed star-forming clumps or complexes are much more massive at high redshifts (e.g., \citealt{Tacconi2010,FosterSchreiber2011,Swinbank2011}), this criterion should only move to higher masses at high-$z$, consistent with our findings. Further, from post-processing cosmological simulations of massive galaxies with well-defined dynamical centers merging at $z<6$, \cite{2019MNRAS.486..101P} also concluded that it was crucial that BHs are already well-anchored to the galaxy centers before and throughout mergers, and that the centers are well-defined and dense enough to avoid tidal disruption, in order for BHs to ``sink.'' They specifically concluded that it was crucial that BHs be embedded either in a dense satellite nucleus or a massive nuclear star cluster. This is essentially identical to our ``solution 2'' above. \cite{2019MNRAS.486..101P} also noted that in the cosmologically simulated galaxies at earlier times, when the universe is $<1\,$Gyr old, even with their most massive ($\sim 10^{5}\,M_{\odot}$) seeds, the model for DF does not help in keeping BHs in the center, as the galaxy is so chaotic that BHs wander no matter the implementation of DF. This is again in good agreement with our conclusion.

A recent study by \cite{Trebitsch2020} provides another excellent illustration of our key conclusions, in a single case-study of a galaxy simulation with ``live'' AGN accretion and feedback. While the authors found that they could produce rapid BH growth by $z\sim 6$, they (1) had to impose a sub-grid DF model with an artificial super-linear density dependence ($\propto \rho^{3}$ at high densities) designed to ``anchor'' BHs into high-density regions (essentially our solution 2, again); (2) still found almost no BH growth until $z\lesssim 8$, after the galaxy reaches $M_{\ast} \gg 10^{9}\,M_{\odot}$ and forms a dense, strongly-peaked and well-defined central ``proto-bulge'' structure, very much like the late-time-forming structures we argue are necessary for BH capture and retention; and (3) still only reach peak luminosities $\ll 10^{43}\,{\rm erg\,s^{-1}}$ in X-rays, about a factor of $\sim 10^{3}-10^{4}$ less-luminous than the most luminous QSOs observed at these redshifts \citep{Shen2019}, which makes them still challenging to form.

There are some recent studies which might appear to be in contrast to our results at first glance. For instance, \cite{tremmel2018b} has shown that host galaxies could aid SMBHs to shorter sinking timescales, and the {\small ROMULUS} simulations \citep{Tremmel2017,Tremmel2019,Ricarte2019} argue that it is possible to grow massive black holes by intermediate redshifts. But a closer comparison shows these simulations are consistent with all of our key conclusions. In these studies, the BHs are, as the authors note \citep{tremmel2018b}, embedded in nuclear regions of the host galaxy, which are dense enough to avoid tidal disruption and much more massive than the BHs. The nuclear regions, with high ``effective mass'', hence sink as a whole -- again following our  ``solution 2'' above. This is effective because these studies focus on cases where the galaxies are already massive, with unambiguous massive central peaks in their density profiles at relatively low redshift (with $z\lesssim 2-4$, cf. Fig 4 in \citealt{tremmel2018b}). Moreover, in e.g.\ {\small ROMULUS}, the simulations have an effective seed mass $\sim 10^{6}-10^{7}\,M_{\odot}$\footnote{The authors note that their seed criterion often produces multiple seeds in the same kernel which are instantly merged, producing a range of effective initial seed masses.}, close to our sinking mass threshold in a smooth galaxy. These demonstrate that, given enough time and a pre-existing massive density peak to ``anchor'' a SMBH, BHs can indeed grow following e.g. our solution 2 as speculated above. Our focus here is essentially on how the ``initial conditions'' of these simulations (at earlier times and smaller mass and spatial scales) could arise. We focus on galaxies at much higher redshifts, where those dense central regions either do not exist, or have formed relatively recently (e.g. $z<9$) and one wishes to form an extremely-massive SMBH by $z>7$, significantly shortening the available time for BH growth, especially from extremely low-mass seeds.

\section{Conclusions}
\label{sec:conclusions}

In this study, we explore high-resolution cosmological galaxy formation simulations to understand the dynamics of BH seeds at high-$z$ and their implications for SMBH formation and growth. Our simulations and semi-analytic DF calculations show that BH seeds cannot efficiently ``sink'' to galaxy centers and/or be retained at high redshifts unless they are extremely massive already, $M > 10^{8}\,M_{\odot}$, i.e.\ already SMBHs. We show that this threshold is at least an order-of-magnitude higher than what one would expect in a spherically-symmetric smooth galaxy potential, as commonly adopted in analytic or older simulation calculations which could not resolve the complex, clumpy, time-dependent sub-structure of these galaxies. For smoother galaxies, this mass threshold reduces to $10^7\,M_\odot$, which does not change the key conclusion. 

We therefore join the growing number of recent studies by different groups which have reached similar conclusions \citep[see e.g.][]{daa:BHs.on.FIRE,2017MNRAS.469..295B,2018ApJ...857L..22T,2019MNRAS.486..101P,Bellovary2019,Barausse2020,2020MNRAS.495L..12B}. All of these studies, like ours, have concluded that this ``sinking problem'' for BH seeds may, in fact, be even more challenging than even other well-known challenges for explaining the formation and growth of the first SMBHs with masses $\gg10^{9}\,M_{\odot}$ in galaxy centers at redshifts $z > 7$. Our contributions to extending this previous work include: (a) studying fully-cosmological simulations with higher resolution, a broader range of redshifts, a much broader spectrum of BH seed masses, and different (sometimes more detailed) explicit models for stellar feedback; (b) comparing direct cosmological simulations which only resolved N-body dynamics to semi-analytic post-processing models for DF, to verify that these conclusions are robust; and (c) extending our comparisons to the ``test particle limit'' by treating all stars as possible BH seeds.

Like these other studies, we qualitatively conclude that the chaotic, rapidly time-evolving, clumpy, bursty/dynamical nature of high-redshift galaxies, coupled to the very short Hubble times ($\lesssim 1\,$Gyr) make it nearly impossible for any lower-mass seeds to efficiently ``migrate'' from $\gtrsim 1$\,kpc scales to galaxy centers, and is far more likely to eject seeds than to retain them. Like these authors concluded, the clumpy, bursty nature of the ISM is crucial for these conclusions: so this can only been seen in simulations which resolve the cold phases of the ISM and explicitly model stellar feedback. It is also worth noting that for low-mass galaxies (the progenitors where, in most models, seeds are supposed to have formed), even at $z\sim 0$, clumpiness and burstiness are ubiquitous, and it is not simply a question of dynamical perturbations but even more basically of the fact that {\em dwarf and high-redshift galaxies do not have well-defined dynamical centers} to which anything {\em could} ``sink.'' This is true even for well-evolved galaxies such as the LMC today. 

In fact, we show that even the extremely massive BHs ($\gtrsim 10^{8}\,M_{\odot}$) which do ``sink'' actually do not sink to the same location at sub-kpc scales, where their migration stalls. This has potentially profound implications for LISA detections of SMBH-SMBH mergers in high-redshift galaxies. Essentially, the ``last parsec problem'' so well-studied in the extremely dense, smooth, well-defined bulges of $z=0$ galaxies (where the Hubble time is long) becomes a ``last kiloparsec problem'' in these galaxies.

Solutions to the ``sinking problem'' for SMBH growth/formation generically fall into one of two categories which we discuss in detail. (1) Either seeds form ``in situ'' when the massive bulge finally forms and creates a deep central potential, or a large number of seeds form so that even the infinitesimally small fraction which have just the right orbital parameters to be ``captured'' by this bulge can exist. In either case, the problem is that we show this deep central potential well does not form until quite ``late,'' at redshift $z\lesssim 9$, from gas and stars which are already highly metal-enriched (metallicities $\gtrsim 0.1\,Z_{\odot}$). This would mean popular speculative BH seed formation channels like Pop III relics or ``direct collapse'' from hyper-massive quasi-stars could not provide the origin of the SMBHs. Moreover, the combination of the fact that this occurs late, and that the stellar IMF is ``normal'' at these metallicities, means that the ``timescale'' problem is much more serious: stellar-relic BHs, if primarily growing by accretion in such massive bulges, must grow from $\sim 10\,M_{\odot}$ to $\gg 10^{9}\,M_{\odot}$ in $\lesssim 200\,$Myr -- requiring sustained highly super-Eddington accretion. Alternatively (2) ``seed'' BHs must have enormous ``effective'' masses to form early and remain ``trapped'' and/or sink efficiently to the growing galaxy center. Of course, BHs ``born'' with $M_{\rm BH} \gg 10^{7}\,M_{\odot}$ would solve this, but only by bypassing any stage that could be called a ``seed'' (moreover, no serious models involving standard-model physics can produce seeds of such large mass). However, models where seeds preferentially form tightly-bound in dense star cluster centers owing to physics not modeled here (for example, runaway stellar mergers in the center of dense, high-$z$ massive star clusters; see \citealt{shi2020}) could (if the cluster is sufficiently dense) have an ``effective'' dynamical mass for our purposes  of roughly the cluster itself, which could reach such large values. This suggests these regions may be promising sites for SMBH seed formation.

In future work, we will explore the role of BH accretion and feedback, and more explicitly consider models where BH seeds form in resolved star clusters, as well as a wider range of galaxy simulations. It is likely that {\em all} of the scenarios above require a sustained period of super-Eddington accretion, so we will also explore whether this requires seed BHs residing (or avoiding) certain regions within high-$z$ galaxies. We have also neglected models where non-standard model physics (e.g.\ dissipative dark matter, primordial BHs) allows for new formation channels and test-body dynamics. We will also explore new applications of our numerical DF approximator, in a variety of other interesting contexts (e.g. pairing of SMBHs in massive galaxy mergers at low redshifts).

\acknowledgments{We thank Zuyi Chen and Alessandro Lupi for useful discussions. Support for LM \&\ PFH was provided by NSF Research Grants 1911233 \&\ 20009234, NSF CAREER grant 1455342, NASA grants 80NSSC18K0562, HST-AR-15800.001-A. DAA acknowledges support by NSF grant AST-2009687 and by the Flatiron Institute, which is supported by the Simons Foundation. CAFG was supported by NSF through grants AST-1715216 and CAREER award AST-1652522; by NASA through grant 17-ATP17-0067; and by a Cottrell Scholar Award and a Scialog Award from the Research Corporation for Science Advancement. Numerical calculations were run on the Caltech compute cluster ``Wheeler,'' allocations FTA-Hopkins supported by the NSF and TACC, and NASA HEC SMD-16-7592.}

\section*{Data Availability}
The data and source code supporting the plots within this article are available on reasonable request to the corresponding author.

\bibliography{seed_sink}

\end{document}